\documentclass[11pt,twoside]{article}

\usepackage{asp2006}
\usepackage{epsf}
\usepackage{psfig}
\usepackage{lscape}

\begin{document}
\title{Accretion Events in Binary Systems: AZ Cas and VV Cep}
\author{C. Ga\l{}an,$^1$ M. Miko\l{}ajewski,$^1$ T. Tomov,$^1$ M. Wi\c{e}cek,$^1$ A. Majcher,$^2$
P.~Wychudzki,$^1$ E. \'Swierczy\'nski,$^1$ D. Kolev,$^3$ T. Bro\.zek,$^1$ G.~Maciejewski,$^1$
S. Zo\l{}a,$^{4,5}$ M. Kurpi\'nska--Winiarska,$^5$ M. Winiarski,$^5$ W. Og\l{}oza,$^4$
M. Dro\.zd\.z,$^4$ and J. Krzesi\'nski$^4$
}

\affil{$^1$Nicolaus Copernicus University, Torun, Poland\\
$^2$Soltan Institute for Nuclear Studies, Warsaw, Poland\\
$^3$NAO Rozhen, Institute of Astronomy, Smolyan, Bulgaria\\
$^4$Mt. Suhora Observatory, Pedagogical University, Cracow, Poland\\
$^5$Astronomical Observatory, Jagiellonian University, Cracow, Poland
}

\begin{abstract}
The sudden lengthening of orbital period of VV Cep eclipsing binary by
about $1\%$ was observed in the last epoch. The mass transfer and/or mass
loss are most possible explanations of this event. The photometric
behaviour of AZ Cas, the cousin of VV Cep, suggests that the accretion
can occur and could be important in this system, too.
\end{abstract}

\section*{Introduction}
AZ Cas and VV Cep are eclipsing binary stars with one of the longest orbital periods
($9.3$ and $20.3$ years, respectively). They both belong to VV Cep type--systems, consisting
of a M or late K supergiant and an early B star, and have spectra exhibiting strong Balmer
and $[FeII]$ emission lines \citep{Cow1969}. The presence of an intensive mass loss gives us
an opportunity to study the accretion processes. The most recent, 1997 eclipse of VV Cep
occured by about $1\%$ of the orbital period later than predicted \citep{Gra1999}, perhaps
as a result of mass loss and/or mass transfer. The behaviour of both systems seems to have
the same cause, connected with strong interactions between the components close to periastron
phase and it can influence on times of minima due to changes in the orbital parameters.

\section*{Observations of the last eclipses}
Multicolor photometry of the last AZ Cas eclipse was obtained with a 60 cm reflector at
Piwnice Observatory (Poland). Our \textit{UBVR} light curves are presented in the Figure~1 (top)
phased together with photometric data from other observatories. Our \textit{BVR} photometry after
the eclipse shows an increase of the brightness by a few $\%$ of magnitude during $\sim250$ days.
This effect was commented by \citet{Tem1980}, and as a first approximation it can be interpreted
as effect of tidal distortion of the cool component surface around the periastron.
Nevertheless, our observations are made also in phases which have never been covered with
multicolor photometry before. For example, the long atmospheric eclipse in \textit{U} bandpass
(up to $\sim0.1$ phase) seems to indicate, that the cool component is surrounded by an extended
envelope formed by its own stellar wind. The hot component could accrete gas from the wind and the
matter transfered through the L$_1$ point by the cool giant Roche lobe overflow near the periastron.
In the out of eclipse data (the phase range from $\sim0.25$ to $\sim0.45$) we have observed very
interesting variability with smallest amplitude in \textit{V} bandpass, and much stronger
in the \textit{U} and in \textit{U-B} color index. The color indices evolution (Figure~1--inner~box)
can not be explained at this time, and we can only assume, that this is connected with accretion of
material which was lost by supergiant near the periastron passage and with cooling of supergiant
atmosphere.

\begin{figure}[!ht]
\plotone{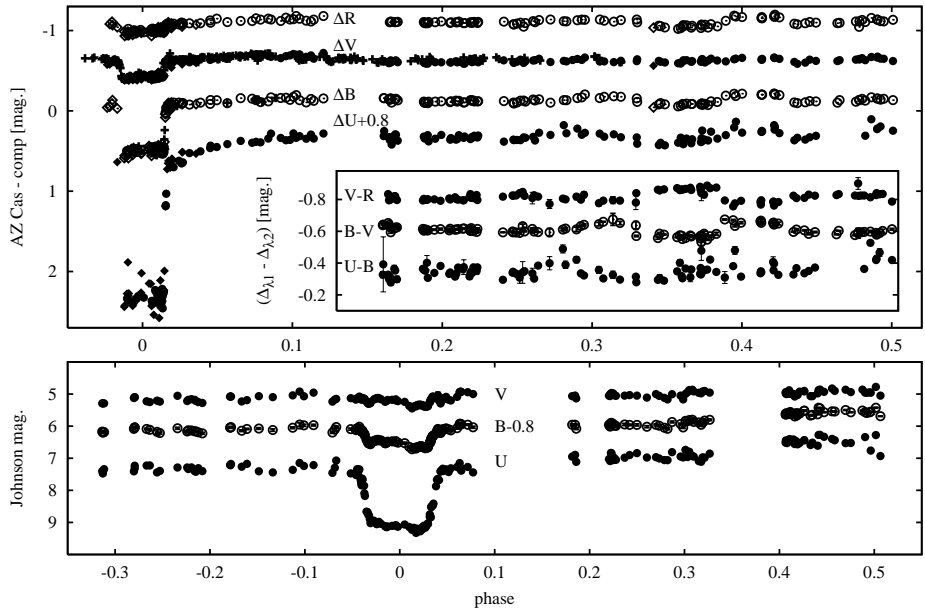}
\caption{\small \textit{UBVR} light curves of AZ Cas (top). Circles: Piwnice Observatory data points.
Diamonds: Suhora and Cracow measurements \citep{Mik2004}. Croses: \citet{Lar1959} and
\citet{Tem1980} data. Changes of the \textit{U-B}, \textit{B-V} and \textit{V-R} color indices
of AZ Cas (top--inner~box). \textit{UBV} light curves of VV Cep (bottom).}
\end{figure}

The VV Cep system has been observed in Piwnice Observatory for $\sim17$ years. The collected \textit{UBV}
light curves are presented in the Figure~1 (bottom pannel). Especially strong variations in the \textit{U}
bandpass can be attributed to mass accretion on the hot component, as they disappear during the eclipse.
The delay of the last eclipse event could occur, as the result of changes in orbital parameters
(perhaps $e$ and $\omega$) due to interaction and mass loss/transfer near the periastron passage in 
the previous epoch.

\section*{AZ Cas spectral type changes}
\citet{Men1975} suggested that the spectral type of the cool component can vary with
the orbital phase. Around the periastron phases, the cool component can look like a K
supergiant (e.g.: K3.8Ib/Iab \citep{Waw1974}) and perhaps even close to F type during
the eclipse (F8Ib \citep{Men1975}). In most cases the cool component is classified as M type
supergiant (e.g.: M0Ib \citep{Lee1970}). We used a Coude spectrograph at the Rozhen Observatory
(Bulgaria) to obtain spectra of AZ Cas in totality in 2003, and echelle spectrograph at the Asiago
Observatory (Italy) to obtain one spectrum near the periastron in 2004. In Figure~2 our spectra are
compared with the spectra of HD~11800 (K5Ib) and HD~12014 (K0Ib) taken from the library of
\citet{Mon1999}. 

\begin{figure}[!ht]
\plotone{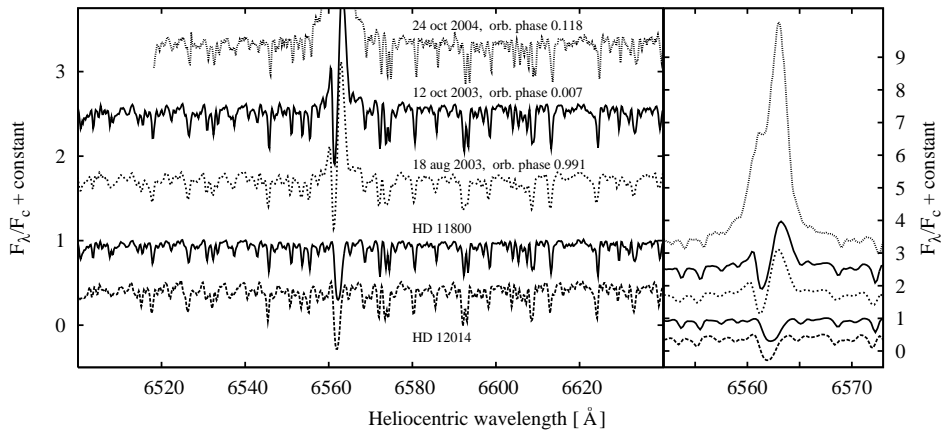}
\caption{\small The Rozhen and Asiago spectra of AZ Cas ($R\sim16000$) compared with
the spectra of HD 11800 (K5Ib) and HD 12014 (K0Ib).}
\end{figure}

In our spectra the cool component looks like an early K type supergiant.
During the totality there is seen $H_{\alpha}$ emission with superimposed absorption
(Figure~2: right) whereas the $H_{\beta}$ emission is absent. $H_{\alpha}$ emission
seems to be especially strong close to the periastron. In Figure~3 (left panel) the \textit{U-B},\\ 
\textit{B-V} color--color diagram is presented. The reddenings for the hot and the cool
components are different and \textit{$E_{B-V}$} have values $0.90\pm0.11$ and $0.64\pm0.22$
respectively, which cause an impression that cool component could be somewhat more blue at
the eclipse, possibly as a result of an excess of radiation in short wavelengths.
During totality, the supergiant looks like K0--2 type star.

\section*{Ellipsoidal and scattering effects in AZ Cas}
A model resuling from the \citet{Wil1971} code, adopting \citet{Cow1977} orbital parameters
$e=0.55$, $\omega=4^{\circ}$, can explain quite well the brightening of the system near the
periastron as a result of the ellipsoidal effect. However there is a visible excess of radiation
near and during the eclipse (Figure~3--right panel), which could be produced by a scattering of
radiation emitted by the hot component in the very extended envelope of the supergiant. Such
effect could also explain the brightening observed in the mid--eclipse phase (see Figure~1).

\begin{figure}[!ht]
\plotone{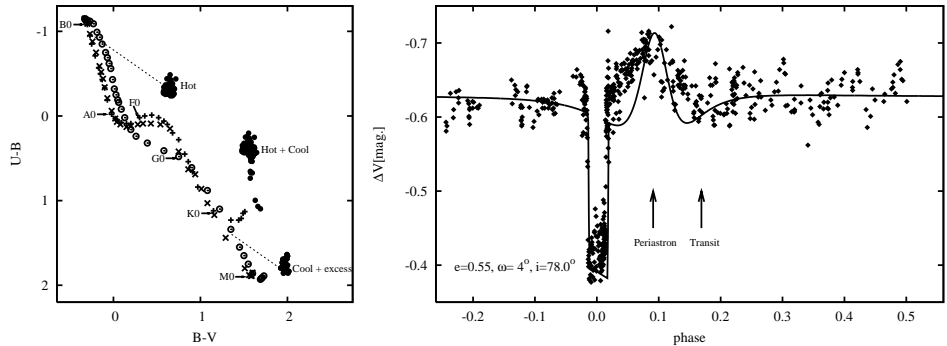}
\caption{\small On the left \textit{U-B}, \textit{B-V} color--color diagram is presented, where the
position of the system during the eclipse (Cool + excess), out of eclipse (Hot + Cool), and only hot
component extracted (Hot) are shown. Strai\v{z}ys calibrations are drawn with symbols
$+,\times,\sun$ for $V,III$ and $I$ luminosity class respectively. On the right the synthetic
\textit{V} light curve is compared with the observational data.}
\end{figure}

\section*{Conclusions}
1. Both systems show flux changes in hot component near and after the periastron passage.\\
2. The cool component is surrounded by an extended envelope, because:\\
a. broad atmospheric eclipse in \textit{U} bandpass is visible,\\
b. the color of the cool component is more blue during the eclipse (scattering of the hot
component light in the supergiant envelope),\\
c. the absorption component in $H_{\alpha}$ appears during the eclipse.\\
3. Mass loss/transfer leads to changes of the orbital parameters ($e$, $T_o$, $\omega$, $i$, $P$)
and not only in the orbital period as it was suggested by \citet{Gra2001}.

\acknowledgements
We are very grateful to K. Czart, S. {Fr\c{a}ckowiak}, M. Hajduk, A. Karska,
J. Osiwa\l{}a, P. Oster, G. Pajdosz, P. {R\'o\.za\'nski}, P. Wirkus,
for their contribution to collection of photometric and spectroscopic data.
This study was supported by MNiSW grant No.~N203~018~32/2338.

\end{document}